\documentclass[12pt,a4paper,aps,preprint,superscriptaddress,nofootinbib]{revtex4-1}
\usepackage[utf8]{inputenc}
\usepackage{graphicx}
\usepackage{amssymb}
\usepackage{textcomp}
\usepackage{amsmath}
\usepackage{tabularx}
\usepackage{bm}
\usepackage{times}
\usepackage{color}
\usepackage{slashed}
\usepackage{multirow}
\usepackage{verbatim}
\usepackage{cancel}
\usepackage{subfigure}
\usepackage{ulem}

\usepackage[colorlinks=true, pdfstartview=FitV, linkcolor=blue, citecolor=blue, urlcolor=blue]{hyperref}
\allowdisplaybreaks[4]

\def\lsim{\mathrel{\raise.3ex\hbox{$<$\kern-.75em\lower1ex\hbox{$\sim$}}}}
\def\gsim{\mathrel{\raise.3ex\hbox{$>$\kern-.75em\lower1ex\hbox{$\sim$}}}}

\newcommand{\eV}{{\rm eV}}
\newcommand{\keV}{{\rm keV}}
\newcommand{\MeV}{{\rm MeV}}
\newcommand{\GeV}{{\rm GeV}}
\newcommand{\TeV}{{\rm TeV}}
\newcommand{\Tr}{{\rm Tr}}

\definecolor{orange}{rgb}{1,0.5,0}

\begin{document}

\title{Probing light axion-like particle via primordial black hole evaporation with gamma-ray observations}

\author{Tong Li}
\email{litong@nankai.edu.cn}
\affiliation{
School of Physics, Nankai University, Tianjin 300071, China
}
\author{Rui-Jia Zhang}
\email{zhangruijia@mail.nankai.edu.cn}
\affiliation{
School of Physics, Nankai University, Tianjin 300071, China
}

\begin{abstract}
The axion-like particle (ALP) and primordial black hole (PBH) are two representative dark matter (DM) candidates as light bosonic DM and macroscopic objects, respectively. In this work, we investigate the gamma-ray production mechanisms induced by PBH evaporation and ALP-photon coupling $g_{a\gamma\gamma}$. The detection of gamma-rays is also explored in future satellite telescopes, including AMEGO, e-ASTROGAM and MAST.
We first propose the evaporation-conversion scenario in which light ALPs are emitted by PBHs and are converted into photons in the presence of magnetic field in the Universe. The second scenario assumes ALPs as dominant DM component in the Milky Way and considers the relativistic electron production from PBH evaporation. The emitted electrons scatter off non-relativistic ALP in DM halo and produce gamma-rays through the ALP-photon coupling.
Using the Fisher forecasting method, we calculate the gamma-ray energy spectra from these two scenarios and derive projected sensitivity for the fraction of DM composed of PBHs $f_{\rm PBH}$ and ALP-photon coupling $g_{a\gamma\gamma}$.
\end{abstract}

\maketitle

\tableofcontents

\section{Introduction}
\label{sec:Intro}

A wealth of astrophysical observations has established robust evidence supporting the existence of Dark Matter (DM). However, both the nature and fundamental properties of DM still lack a complete understanding. In recent years, significant effort has been devoted to the theoretical hypotheses of DM candidates beyond the weakly interacting massive particle (WIMP) and their detection methods in experiments. In general, DM can arise through many distinct mechanisms in
theory and requires rather different experimental strategies and facilities. The range of DM mass may span from sub-micro-eV to TeV scale and even beyond (see a recent review~\cite{Cirelli:2024ssz}).
The axion-like particle (ALP) and primordial black hole (PBH) are two representative DM candidates as light/ultralight bosonic DM and massive macroscopic objects, respectively.

The ALP as a CP-odd pseudo-Nambu-Goldstone boson is the result of the spontaneous breaking of a global $U(1)$ symmetry~\cite{Peccei:1977hh,Peccei:1977ur,Weinberg:1977ma,Wilczek:1977pj} (see a recent review for QCD axion in Ref.~\cite{DiLuzio:2020wdo}). The ALP mass ($m_a$) and its associated symmetry breaking scale (commonly denoted as the decay constant $f_a$) need not be fundamentally related~\cite{Dimopoulos:1979pp,Tye:1981zy,Zhitnitsky:1980tq,Dine:1981rt,Holdom:1982ex,Kaplan:1985dv,Srednicki:1985xd,Flynn:1987rs,Kamionkowski:1992mf,Berezhiani:2000gh,Hsu:2004mf,Hook:2014cda,Alonso-Alvarez:2018irt,Hook:2019qoh}. It can play as cold DM through the misalignment mechanism~\cite{Abbott:1982af,Dine:1982ah,Preskill:1982cy} and the typical mass of ALP making up DM is around micro-eV~\cite{GrillidiCortona:2015jxo} or even down to $\sim 10^{-22}$ eV~\cite{Hu:2000ke,Hui:2016ltb,Niemeyer:2019aqm}.
PBHs may form through gravitational collapse of local density perturbations in the early Universe and potentially serve as macroscopic DM candidates~\cite{Zeldovich:1967lct,Carr:1974nx,Carr:1975qj,Khlopov:2008qy,Barrau:2003xp} (see Ref.~\cite{Carr:2021bzv} for a recent review).
The famous discovery by S.~Hawking tells that PBHs would thermally radiate elementary particles with masses less than the effective temperature~\cite{Hawking:1974rv,Hawking:1975vcx}. The elementary particles emitted from PBHs such as gamma-rays, neutrinos and electrons in the evaporation process suffer from a variety of cosmological constraints~\cite{Carr:2020gox,Laha:2019ssq,Laha:2020ivk,Saha:2021pqf,Ray:2021mxu,Dasgupta:2019cae,Wang:2020uvi,Calabrese:2021zfq,DeRomeri:2021xgy,Ghosh:2021vkt,Capanema:2021hnm,Chao:2021orr,Bernal:2022swt,Huang:2024xap,Wu:2024uxa,Zantedeschi:2024ram,Klipfel:2025jql}. Interestingly, the mechanism and phenomenology of new particles beyond the Standard Model (SM) from the Hawking radiation were also studied in very recent works~\cite{Calabrese:2021src,Li:2022jxo,Calabrese:2022rfa,CDEX:2022dda,Li:2022xqh,Jho:2022wxd,Agashe:2022phd,DeRomeri:2024zqs,Federico:2024fyt,Cheung:2025gdn}.
The emitted particles can acquire energies peaked around a few hundred MeV from the evaporation of PBHs with the asteroid-scale mass $M_{\rm PBH}\gtrsim 10^{15}$ g. The subsequent daughter particles then gain enough kinetic energies to reach the satellite telescopes or terrestrial detectors on Earth.

We once proposed the scenario of
light ALP emittance from the Hawking radiation of PBH with asteroid mass scale (referred as ``PBH ALP'') and the relevant detection in terrestrial detectors~\cite{Li:2022xqh}. Later, Refs.~\cite{Jho:2022wxd,Agashe:2022phd} studied the detection of gamma-rays from the decay of ALP produced by PBH evaporation.
In this work, we suggest alternative gamma-ray production mechanisms from light PBH ALP and explore the detection in future satellite telescopes. The light ALP can be coupled to photons through an extremely weak coupling
\begin{eqnarray}
-{1\over 4}g_{a\gamma\gamma} a F^{\mu\nu} \tilde{F}_{\mu\nu}\;,
\label{eq:ALP-photon}
\end{eqnarray}
where $F_{\mu\nu}$ is the SM electromagnetic field strength and $\tilde{F}_{\mu\nu}$ is its Hodge dual. As a consequence of this ALP-photon coupling, photon-ALP oscillations can happen
in the presence of an external magnetic field~\cite{DeAngelis:2011id}.
After being emitted by PBHs, ALPs propagate through cosmic magnetic fields and can convert into photons which would be probed by gamma-ray satellite experiments.
This scenario is particularly sensitive to ALPs with much smaller masses.
On the other hand, we also consider the scenario of ALP being dominant component of the Milky Way DM halo and the electron production from PBH evaporation. The energetic electrons scatter off non-relativistic ALP in DM halo and emit gamma-rays through the ALP-photon coupling in an inverse Primakoff process~\cite{Dent:2020qev,Goncalves:2025nij}.

The resulting gamma-ray signals from both scenarios could be detectable by the proposed AMEGO~\cite{AMEGO:2019gny,Kierans:2020otl}, e-ASTROGAM~\cite{e-ASTROGAM:2016bph,e-ASTROGAM:2017pxr} and MAST~\cite{Dzhatdoev:2019kay} experiments with specific spectral windows. AMEGO covers the energy range of $150~\keV-5~\MeV$, while e-ASTROGAM and MAST extend to higher energies spanning $100~\MeV$ to $3~\GeV$. AMEGO has the largest effective area in the low-energy range, making it ideal for MeV-scale observations. In contrast, e-ASTROGAM provides an effective area of $\mathcal{O}(10^3)~{\rm cm}^2$, while MAST achieves a significantly larger effective area of $\mathcal{O}(10^5)~{\rm cm}^2$.
We will show the prospect of probing ALP-photon coupling $g_{a\gamma\gamma}$ and the bound on the fraction of DM composed of PBHs $f_{\rm PBH}$ in these three experiments. The calculation of PBH radiation into elementary particles is based on quantum mechanism in curved spacetime. The evolution of quantum vacuum states near the black hole horizon to asymptotic states at infinity produces a net thermal particle flux, characterized by an emission rate following Hawking's prediction~\cite{Hawking:1974rv,Carr:1974nx,Carr:1975qj,Carr:2020gox,Carr:2021bzv,Auffinger:2022khh}. The PBH-sourced ALP flux itself is thus independent of the ALP coupling constant; only the subsequent conversion to photons or scattering process depends on this coupling. We expect the bounds on $g_{a\gamma\gamma}$ and $f_{\rm PBH}$ can be improved from satellite experiments, compared with the current cosmic constraints.

This paper is organized as follows. In Sec.~\ref{sec:PBH}, we overview the mechanism of PBH evaporation and show the flux of emitted ALP or electron from Hawking radiation in galactic and extragalactic regions. The signal event rates in the gamma-ray experiments
are then calculated in Sec.~\ref{sec:Signal}. We discuss two scenarios: the ALP conversion under magnetic field and the scattering of electron with axion DM halo. We then show the prospective detection results of PBH and ALP-photon coupling in Sec.~\ref{sec:Sensitivity}. Our conclusions are drawn in Sec.~\ref{sec:Con}.

\section{PBH evaporation in galactic and extragalactic regions}
\label{sec:PBH}

PBHs emit particles through Hawking radiation, a quantum mechanical process characterized by a thermal emission spectrum with temperature $T_{\rm PBH}$~\cite{Hawking:1975vcx,Page:1976df,Page:1977um,MacGibbon:1990zk,MacGibbon:1991tj}
\begin{eqnarray}
    k_BT_{\rm PBH}=\frac{\hbar c ^3}{8\pi GM_{\rm PBH}}\;,
\end{eqnarray}
where $k_B$ is the Boltzman constant, $G$ is the Newtonian constant of gravitation, and $M_{\rm PBH}$ denotes the PBH mass. All particle species with masses below this temperature threshold can be radiated via the PBH evaporation mechanism.
These primary particles may subsequently undergo decay process and generate secondary particle flux. The general PBH emission rate can be expressed as~\cite{Hawking:1971ei,Page:1976df,Page:1976ki}
\begin{eqnarray}
    \frac{d^2N}{dEdt}=\frac{g}{2\pi}\frac{\Gamma(E,M_{\rm PBH},a^*)}{e^{E/T_{\rm PBH}}-(-1)^{2s}}\;,
\end{eqnarray}
where $E$ is the emitted particle energy, $\Gamma$ is the so-called ``graybody'' factor describing the probability of elementary particles escaping the PBH gravitional well, $a^*$ is the PBH spin, $g$ and $s$ denote the degree of freedom and spin of emitted particle, respectively.
We adopt a monochromatic mass distribution for PBHs.
In this work, considering only primary (direct) Hawking emission, we employ the public code \texttt{BlackHawk\_v2.1}~\cite{Arbey:2019mbc,Arbey:2021mbl} to generate the above number density of emitted particle.

\begin{figure}
\centering
\includegraphics[width=0.49\linewidth]{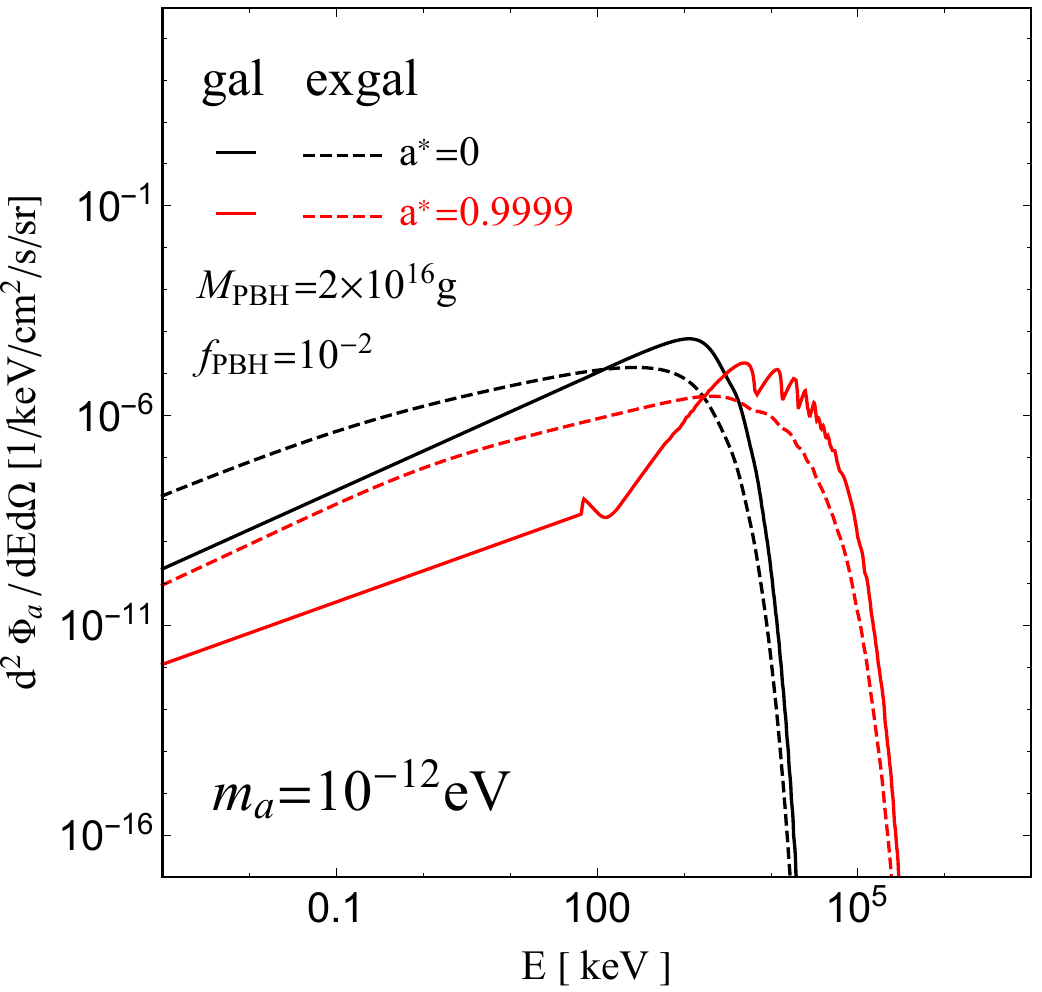}
\includegraphics[width=0.49\linewidth]{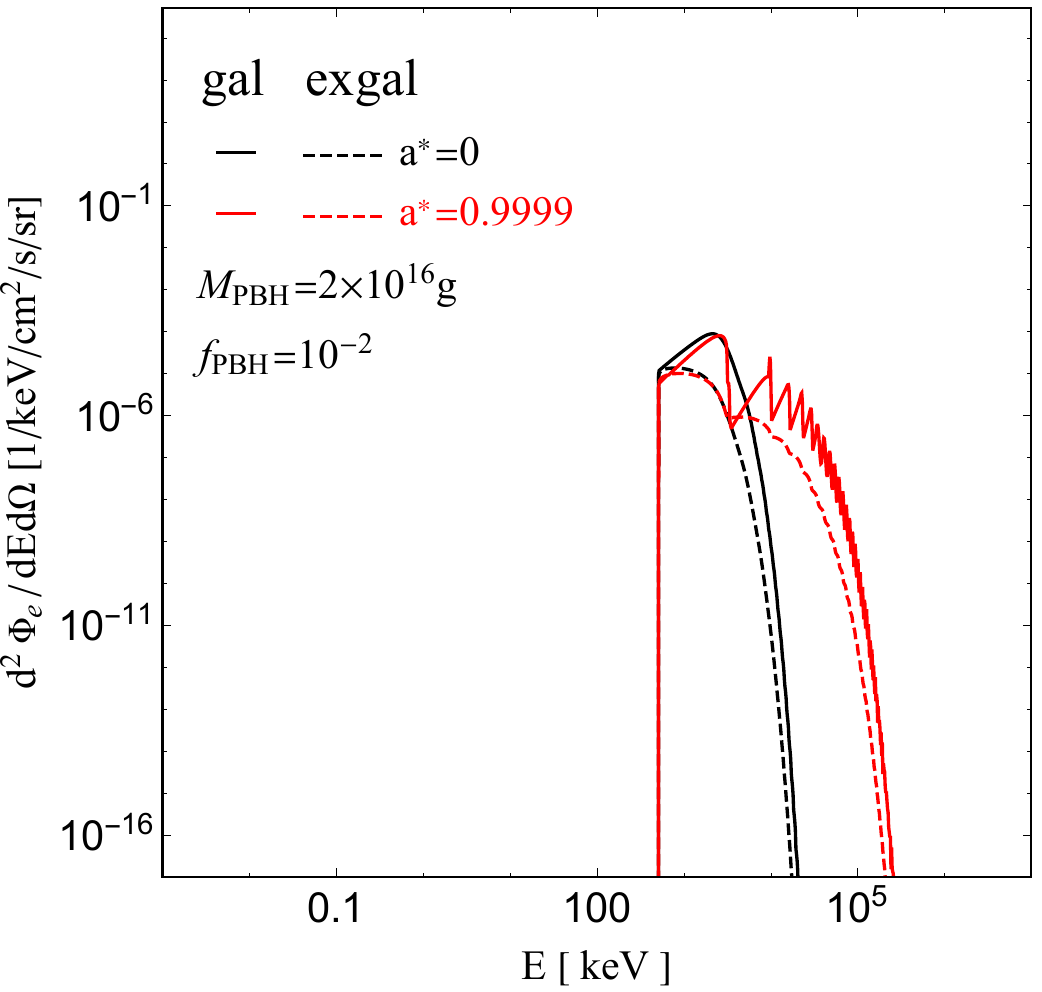}
\caption{The spectrum of ALPs with $m_a=10^{-12}~{\rm eV}$ (left) and electrons (right) from PBH evaporation, assuming $M_{\rm PBH}=2\times 10^{16}$ g and $f_{\rm PBH}=10^{-2}$. Both galactic (solid lines) and extragalactic (dashed lines) PBH contributions are included with $a^\ast=0$ (black) or $a^\ast=0.9999$ (red).
}
\label{fig:PBH_Spectrum}
\end{figure}

The particle flux emitted from PBH can be divided into the
contributions of the PBHs in galactic halo and the extragalactic PBHs
\begin{equation}
\frac{d^2\Phi}{dEd\Omega}=\frac{d^2\Phi_{\rm{gal}}}{dEd\Omega}+\frac{d^2\Phi_{\rm{exgal}}}{dEd\Omega}\;,
\end{equation}
where $\Phi_{\rm gal}$ and $\Phi_{\rm exgal}$ correspond to the flux from galactic (gal) PBHs and extragalactic (exgal) PBHs, respectively, and $\Omega$ is the solid angle.
The evaporation spectrum from PBHs within the Milky Way is given by
\begin{eqnarray}
\frac{d^2\Phi_{\rm gal}}{dEd\Omega}=\frac{f_{\rm PBH}}{4\pi M_{\rm PBH}}\frac{d^2N}{dEdt}\int_{\rm RoI}\frac{d\Omega_s}{\Delta\Omega}\int_0^{s_{\rm max}}ds\rho[r(s,l,b)]\;,
\end{eqnarray}
where $f_{\rm PBH}$ denotes the fraction of DM composed of PBHs, and $\Delta\Omega$ is the integral to solid angle within the Region of Interest (RoI).
We implement the Navarro-Frenk-White (NFW) DM profile~\cite{Navarro:1996gj}
\begin{eqnarray}
    \rho(r)=\rho_\odot\left(\frac{r}{r_\odot}\right)^{-\gamma}\left(\frac{1+r_\odot/r_s}{1+r/r_s}\right)^{3-\gamma}~~{\rm with}~\gamma=1\;,
\end{eqnarray}
where $\rho_\odot=0.4~\GeV/{\rm cm}^3$ is the local DM density~\cite{Salucci:2010qr}, $r_\odot=8.5$ kpc is the distance between the Sun and the Galactic center~\cite{Yuan:2017ozr}, and $r_s=20$ kpc is the radius of the galacitc diffusion disk. The flux of PBHs within the Milky Way requires integrating this DM density distribution along the line-of-sight distance $s$ and over the galactic coordinates $(l,b)$ within the RoI. Here, the galactocentric distance $r$ becomes a function of $s,l$ and $b$: $r=(r_\odot^2+s^2-2sr_\odot\cos l\cos b)^{1/2}$. The differential solid angle element is $d\Omega_s=dl db \cos b$. The maximum line-of-sight distance $s_{\rm max}$ is determined by
\begin{eqnarray}
    s_{\rm max}=r_\odot\cos b\cos l+\sqrt{r_{\rm max}^2-r_\odot^2(1-\cos^2b\cos^2l)}\;,
\end{eqnarray}
with the DM halo radius $r_{\rm max}=200$ kpc~\cite{Wang:2020uvi}. Note that what we calculate here is the \textit{differential flux per unit solid angle}, which is essentially an angular average of a line-of-sight integral, rather than a volume integration. The integration variable $s$ denotes the path length along a fixed direction $(l, b)$, and the integral physically represents the accumulation of the source density (in this case, the PBH emitted particle flux) along that direction. For extragalactic PBHs at cosmologcial distance, we must account for redshift effects and modify the particle emission spectrum as
\begin{eqnarray}
    \frac{d^2\Phi_{\rm exgal}}{dEd\Omega}=\frac{f_{\rm PBH}\rho_{\rm DM}}{4\pi M_{\rm PBH}}\int_{z=0}^\infty\frac{dz}{H(z)}\frac{d^2N}{dEdt}\Bigg\vert_{E\to E_z}\;,
\end{eqnarray}
where the average DM density of the Universe at the present epoch is taken as $\rho_{\rm DM}=2.35\times 10^{-30}~{\rm g}/{\rm cm}^3$ which is determined by Planck 2018 results \cite{Planck:2018vyg}. $E_z=\sqrt{(E^2-m^2)(1+z(t))^2+m^2}$ denotes the energy at the source that is related to the energy $E$ in the observer's frame by the redshift $z(t)$ for a massive particle. $H(z)=H_0\sqrt{\Omega_\Lambda+\Omega_m(1+z)^3+\Omega_r(1+z)^4}$ is the Hubble expansion rate at the redshift $z$ in which $\Omega_\Lambda, \Omega_m$ and $\Omega_r$ are current dark-energy, matter and radiation densities of the Universe, respectively.
Fig.~\ref{fig:PBH_Spectrum} shows the spectrum of ALPs with $m_a=10^{-12}~{\rm eV}$ (left) and electrons (right) from PBH evaporation, assuming $M_{\rm PBH}=2\times 10^{16}$ g and $f_{\rm PBH}=10^{-2}$. Both galactic (solid lines) and extragalactic (dashed lines) PBH contributions are included with $a^\ast=0$ (black) or $a^\ast=0.9999$ (red, limited by the package \texttt{BlackHawk\_v2.1}). The flux of ALP with a small mass extends to much lower energy region, compared with that of electron. A larger PBH spin parameter $a^\ast$ enhances the emission spectrum beyond $\sim 100~{\rm MeV}$.

\section{Gamma-ray signals from PBH evaporation and ALP-photon interaction}
\label{sec:Signal}

When PBHs serve as source of ALPs, the produced ALPs propagating through the galaxies can be efficiently converted into photons via astrophysical magnetic fields~\cite{Raffelt:1987im}.
These gamma-rays, upon reaching the solar system, may be detected by space-based telescope experiments. This gamma-ray production mechanism is termed the ``evaporation-conversion process'' in this work.
Alternatively, if ALPs constitute the local DM halo, PBH electrons may undergo scattering with the ALPs during their propagation through the Milky Way, and also produce gamma-ray signal via the ALP-photon coupling $g_{a\gamma\gamma}$~\cite{Goncalves:2025nij}. This mechnism is hereafter referred to as ``evaporation-scattering process''.

In this section, we will discuss these two processes in details and show the corresponding gamma-ray signal spectra.

\subsection{Oscillation of PBH emitted ALP under magnetic field}

During propagation, the ALPs interact with ambient magnetic fields via the ALP-photon interaction term in Eq.~(\ref{eq:ALP-photon}). Under the resonant condition, this coupling induces coherent ALP-photon oscillations, thereby converting ALPs into detectable photons. A commonly used expression for the ALP-photon conversion probability is given by ~\cite{Raffelt:1987im,Mirizzi:2006zy,Galanti:2022ijh} \begin{equation}
P_{a\to\gamma}=(\Delta_{a\gamma}R)^2\frac{\sin^2(\Delta_{{\rm osc}}R/2)}{(\Delta_{{\rm osc}}R/2)^2}\;,
\label{eq:P_agamma}
\end{equation}
where $R$ is the distance traveled by the ALP, and $\Delta_{{\rm osc}}$ is the oscillation wave number caused by the ALP-photon interaction, which is defined as $\Delta_{{\rm osc}}\equiv\sqrt{(\Delta_a-\Delta_{{\rm pl}})^2+4\Delta_{a\gamma}^2}$. Since the ALPs considered here are ultralight with masses much smaller than eV, they satisfy the extremely relativistic approximation condition $E_a\gg m_a$. In this case, the three wave numbers $\Delta_{a}, \Delta_{a\gamma}, \Delta_{{\rm pl}}$ can be defined as
\begin{eqnarray}
    \Delta_a \equiv -\frac{m_a^2}{2E_a},~\Delta_{a\gamma}\equiv \frac{g_{a\gamma\gamma}B}{2},~\Delta_{{\rm pl}}\equiv -\frac{\omega_{{\rm pl}}^2}{2E_a}\;,
\end{eqnarray}
where $\omega_{{\rm pl}}$ is the effective plasma frequency of photons and is related to the free electrons density in medium with $\omega_{\rm pl}=\sqrt{4\pi\alpha n_e/m_e}\simeq 4\times 10^{-11}~\eV\sqrt{n_e/{\rm cm}^{-3}}$.

However, the actual magnetic field environments in the Universe are not homogeneous, but instead exhibit far more complex structures. The magnetic field in the Milky Way is known to have a complex, multi-scale structure~\cite{Pshirkov:2011um,Jansson:2012pc}. While in galaxy clusters, observations clearly demonstrate that the field exhibits turbulent behavior~\cite{1993ApJ416554T,Ensslin:2003ez,Vogt:2003su}. Given these complexities, the simplified oscillation probability described in Eq.~\eqref{eq:P_agamma} fails to accurately capture the true ALP-photon conversion probability in such realistic astrophysical environments. Therefore, we will subsequently employ a rigorous propagation matrix method to separately calculate the conversion probabilities of PBH ALPs traversing galaxy clusters and the Milky Way.

For a monochromatic photon or ALP beam propagates along $x_3$ axis~\footnote{Regarding the ALP-photon conversion issue, the $x_3$ axis is aligned with the photon propagation direction, while $x_1,x_2$ axes form the transverse plane perpendicular to the propagation direction.} within a cold plasma, the propagation can be described in terms of a Schr\"{o}dinger-like equation~\cite{Raffelt:1987im}
\begin{eqnarray}
    \left(i\frac{d}{dx_3}+E+\mathcal{M}\right)\left(\begin{matrix}
        A_1(x_3)\\A_2(x_3)\\a(x_3)
    \end{matrix}\right)=0\;,
\end{eqnarray}
where $A_1(x_3)$ and $A_2(x_3)$ denote the photon amplitudes with polarization along the $x_1$ and $x_3$ axis, respectively, while $a(x_3)$ is the ALP field strength. $\mathcal{M}$ is referred to ``ALP-photon mixing matrix'' containing the information about ALP-photon interaction.

In galaxy clusters, the magnetic field is not uniform but consists of multiple coherent regions, each with a randomly oriented direction. This implies that the magnetic field exhibits a distinct turbulent structure, where the field direction shows strong randomness across different locations.
According to Ref.~\cite{Horns:2012kw}, the turbulent magnetic field within galaxy clusters can be modeled as multiple coherent domains with fixed strength but random orientations. In each domain, the magnetic field strength is approximately constant, and the direction remains unchanged. However, when transiting to an adjacent domain, the field direction abruptly changes in a random manner. Since the magnetic field orientations are entirely random, the mixing matrix $\mathcal{M}$ is a general three-dimensional matrix
\begin{eqnarray}
\mathcal{M}_k=\left(\begin{matrix}
        \Delta_{xx}&\Delta_{xy}&\Delta_{a\gamma}\sin\psi_k\\
        \Delta_{yx}&\Delta_{yy}&\Delta_{a\gamma}\cos\psi_k\\
        \Delta_{a\gamma}\sin\psi_k&\Delta_{a\gamma}\cos\psi_k&\Delta_a
\end{matrix}\right)\;,
\end{eqnarray}
where the subscript $k$ distinguishes different coherent domains, while $\psi_k$ represents the random magnetic field orientation within each domain.
The matrix elements are given by
$\Delta_{xx}=\Delta_\parallel\sin^2\psi_k+\Delta_\bot\cos^2\psi_k$, $\Delta_{yy}=\Delta_\parallel\cos^2\psi_k+\Delta_\bot\sin^2\psi_k$, $\Delta_{xy}=\Delta_{yx}=(\Delta_\parallel-\Delta_\bot)\sin\psi_k\cos\psi_k$, where $\Delta_\parallel=\Delta_{\rm pl}+7/2\Delta_{\rm QED}$ and $\Delta_\bot=\Delta_{\rm pl}+2\Delta_{\rm QED}$ are two wavenumbers.
For the considered energy window of $\mathcal{O}(1-100)~{\rm MeV}$ in our case, the vacuum polarization effect $\Delta_{\rm QED}$ can be neglected. Consequently, we have $\Delta_\parallel = \Delta_\perp = \Delta_{\rm pl}$ and the matrix element can be further simplied to $\Delta_{xx} = \Delta_{yy} =\Delta_{\rm pl}$, $\Delta_{xy}=\Delta_{yx}=0$.
In this work, we adopt characteristic paramters to model the cluster environment. Specifically, we assume a cluster radius of $R=1~{\rm Mpc}$ and a turbulent magnetic field coherence length of $l_c=10~{\rm kpc}$~\cite{Govoni:2004as}. They result in $k=R/l_c=100$ domains along the propagation path. Past studies indicate that the magnetic field strength in galaxy clusters follows the radial profile of the thermal electron distribution $n_e(r)$, which can be expressed as~\cite{Meyer:2013pny}
\begin{eqnarray}
    B_{\rm cluster}(r)=B^0_{\rm cluster}(n_e(r)/n_e^0)^\eta\;,~~n_e(r)=n_e^0(1+r/r_{\rm core})^{-3\beta/2}\;,
\end{eqnarray}
where $\eta$ typically ranges between $0.5$ and $1$. In our analysis, we adopt $\eta=0.5,~\beta=2/3$, and for the most massive clusters, the central magnetic field and electron number density can reach up to $B^0_{\rm cluster}=10~\mu{\rm G},~n_e^0=10^{-2}~{\rm cm}^{-3}$ with the typical core radius $r_\text{core}=100~\text{kpc}$~\cite{Vikhlinin:2005mp}. Since the magnetic $B(r)$ depends on radius, we determine the magnetic field value of each domian by performing an average over the radial integral.

Given that the ALPs originate from PBH Hawking radiation, their momenta are assumed to be isotropically distributed and their spin is zero. Therefore, the initial density matrix is taken as $\rho_0={\rm diag}(0,0,1)$. The final photon state consists of two linearly polarized pure states, corresponding to $\rho_{\rm final}=\rho_{11}+\rho_{22}={\rm diag}(1,0,0)+{\rm diag}(0,1,0)$. As a result, the conversion probability is given by
\begin{eqnarray}
    P_{a\to\gamma}^{\rm exgal}=\Tr(\rho_{\rm final}\mathcal{T}(\psi_k,...,\psi_1)\rho_0\mathcal{T}^\dagger(\psi_k,...,\psi_1))\;,~~\mathcal{T}(\psi_k,...,\psi_1)=\prod_{k=1}\mathcal{T}_k=\prod_{k=1}e^{i\mathcal{M}_kl_c}\;.
\end{eqnarray}
To fully describe the impact of the turbulent magnetic field, we model the angle $\psi$ as a uniformly distributed random variable in the interval $[0,2\pi)$, independently sampled in each domain. Assuming that PBHs are located at the center of a galaxy cluster, a total of 1000 Monte Carlo realizations are performed, each involving a full-domain propagation with a randomly generated set of angles $\{\psi_k\}_{1\leq k\leq100}$. The median of the resulting conversion probabilities is then adopted as the final statistical result.

\begin{figure}
\centering
\includegraphics[width=0.49\linewidth]{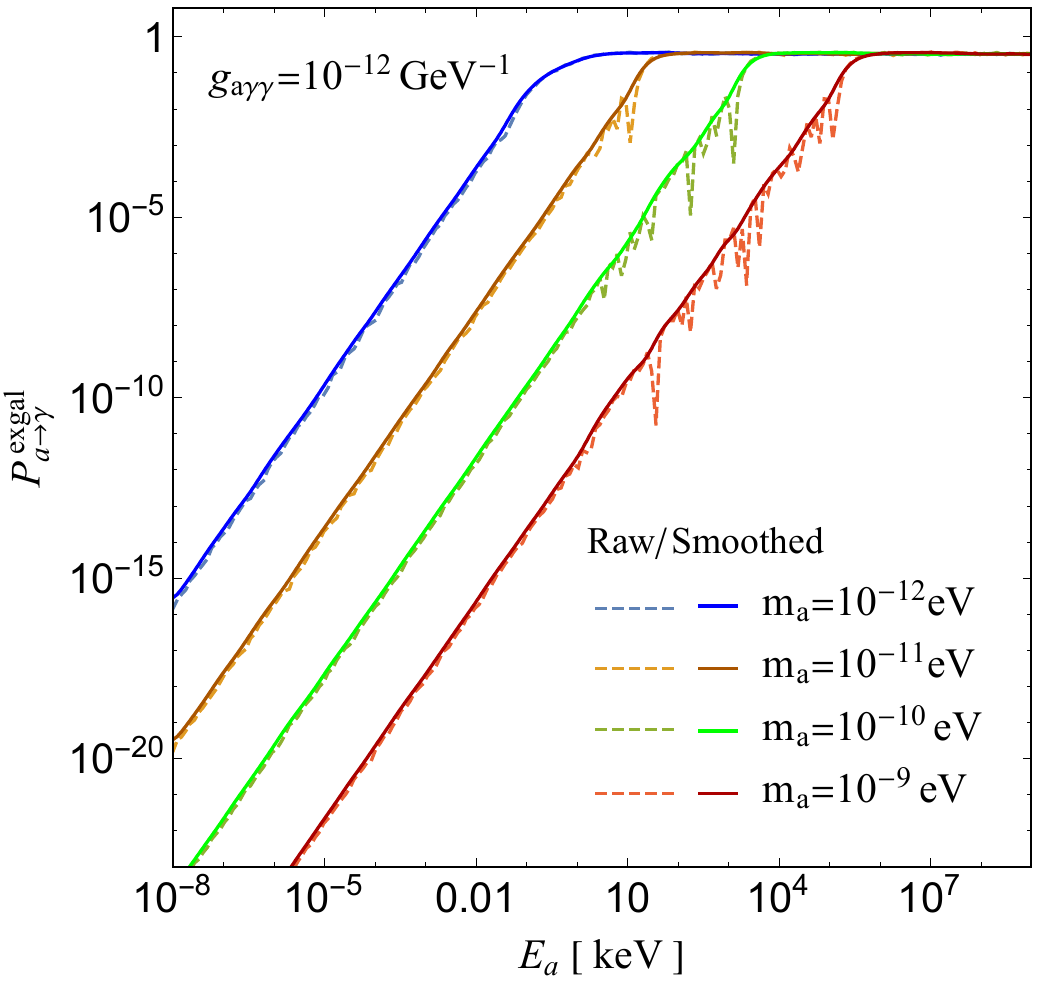}
\includegraphics[width=0.49\linewidth]{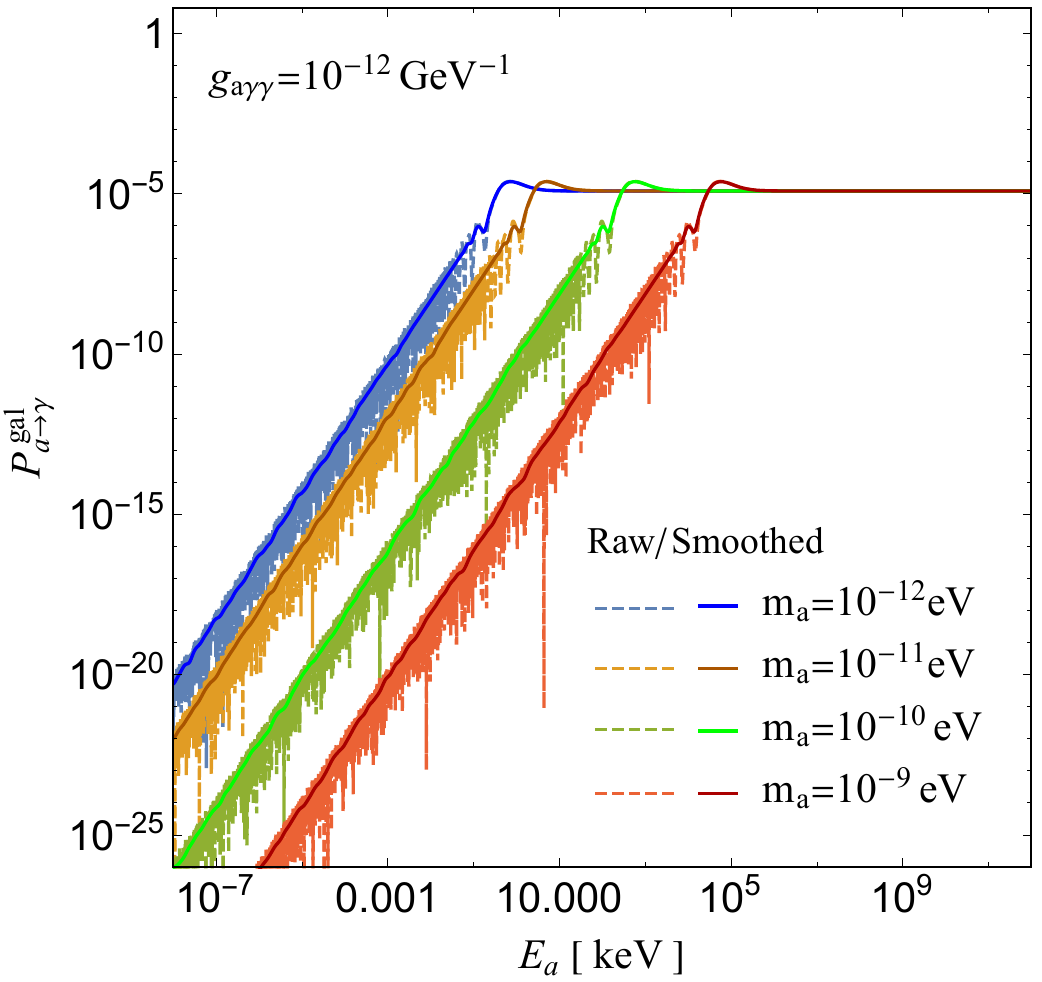}
\caption{Left: Probability of ALP conversion into photon $P_{a\to \gamma}^{\rm exgal}$ within the galaxy clusters. ALP mass is chosen as $m_a=10^{-12}$ eV (blue), $10^{-11}$ eV (orange), $10^{-10}$ eV (green) or $10^{-9}$ eV (red) and ALP-photon coupling is fixed as $g_{a\gamma\gamma}=10^{-12}~\GeV^{-1}$.
Right: Probability of ALP conversion in the Milky Way $P_{a\to \gamma}^{\rm gal}$, calculated for a fixed propagation path from the Galactic Center to the Solar System, with the $g_{a\gamma\gamma}$ and $m_a$ taking the same values as in the galaxy cluster case. We highlight the smoothed curves obtained through Gaussian filtering with solid lines and dark colors, which are then used for our numerical calculations.
}
\label{fig:conversion_probability}
\end{figure}

In the left panel of Fig.~\ref{fig:conversion_probability}, by fixing the coupling $g_{a\gamma\gamma}=10^{-12}~\GeV^{-1}$ and considering four choices of ALP mass $m_a=10^{-12}~\eV$, $10^{-11}~\eV$, $10^{-10}~\eV$, $10^{-9}~\eV$, we show the resultant conversion probability for different ALP energies within the galaxy clusters (dashed lines). One can see that the conversion probability exhibits an oscillatory behavior within a certain region. With increasing energy, the conversion probability becomes stable. To minimize the unstability in numerical calculation caused by the oscillation in probability, we smooth the results by Gaussian filtering as shown by the solid lines and use the smoothed probability for our analysis in the following.

The probability of ALP-photon conversion in the Milky Way requires careful consideration of the Galactic magnetic field structure.
We adopt the ``JF12 model''~\cite{Jansson:2012pc} from the two most commonly used galactic magnetic field models (the other being the Pshikov model from Ref.~\cite{Pshirkov:2011um}). The JF12 model's regular field comprises disk, halo, and X-field components. Here we implement two simplifications as follows. First, by treating the Milky Way as a two-dimensional disk ($z = 0$), we naturally exclude the halo and X-field components and reduce the magnetic field to purely planar configurations along the $\hat{\phi}$ and $\hat{r}$ directions, corresponding to the molecular ring and spiral arm components of the disk field, respectively. Second, since galactic PBH distributions inherit their profiles from DM models that predict strong central concentration, we accordingly consider a single propagation path extending from the Galactic Center to the Solar System, with a characteristic length of 8.2 kpc~\cite{Gravity:2019nxk}.

Along this propagation path, the magnetic field strength falls off as $1/r$. We therefore employ a cell-like approach by dividing the path into eight segments corresponding to the number of spiral arms. Within each domain, both the field strength and orientation are treated as approximately uniform. This allows us to simplify the calculation by rotating the angle $\psi$ reduces to zero.
This rotation makes ALPs only couple to the photon polarization state along the $x_2$-direction, while leaving the $x_1$-polarized photon state completely decoupled. The propagation matrix $\mathcal{M}$ consequently simplifies to
\begin{eqnarray}
    \mathcal{M}_0=\left(\begin{matrix}
        \Delta_\bot&0&0\\
        0&\Delta_\parallel&\Delta_{a\gamma}\\
        0&\Delta_{a\gamma}&\Delta_a
    \end{matrix}\right)\;.
\end{eqnarray}
We then follow Ref.~\cite{Jansson:2012pc} to calculate the averaged magnetic field within each domain, adopting a mean thermal electron density of $n_e=10^{-2}~{\rm cm}^{-3}$ in the Milky Way~\cite{Beck:2008ty,Cordes:2002wz}. Following the method employed for galaxy clusters, the conversion probability within the Milky Way $P_{a\to\gamma}^{\rm gal}$ has been calculated and presented in the right panel of Fig.~\ref{fig:conversion_probability}.
In addition, to minimize the unstability caused by the oscillation in probability, we also smooth the results by Gaussian filtering.

\begin{figure}
    \centering
    \includegraphics[width=0.7\linewidth]{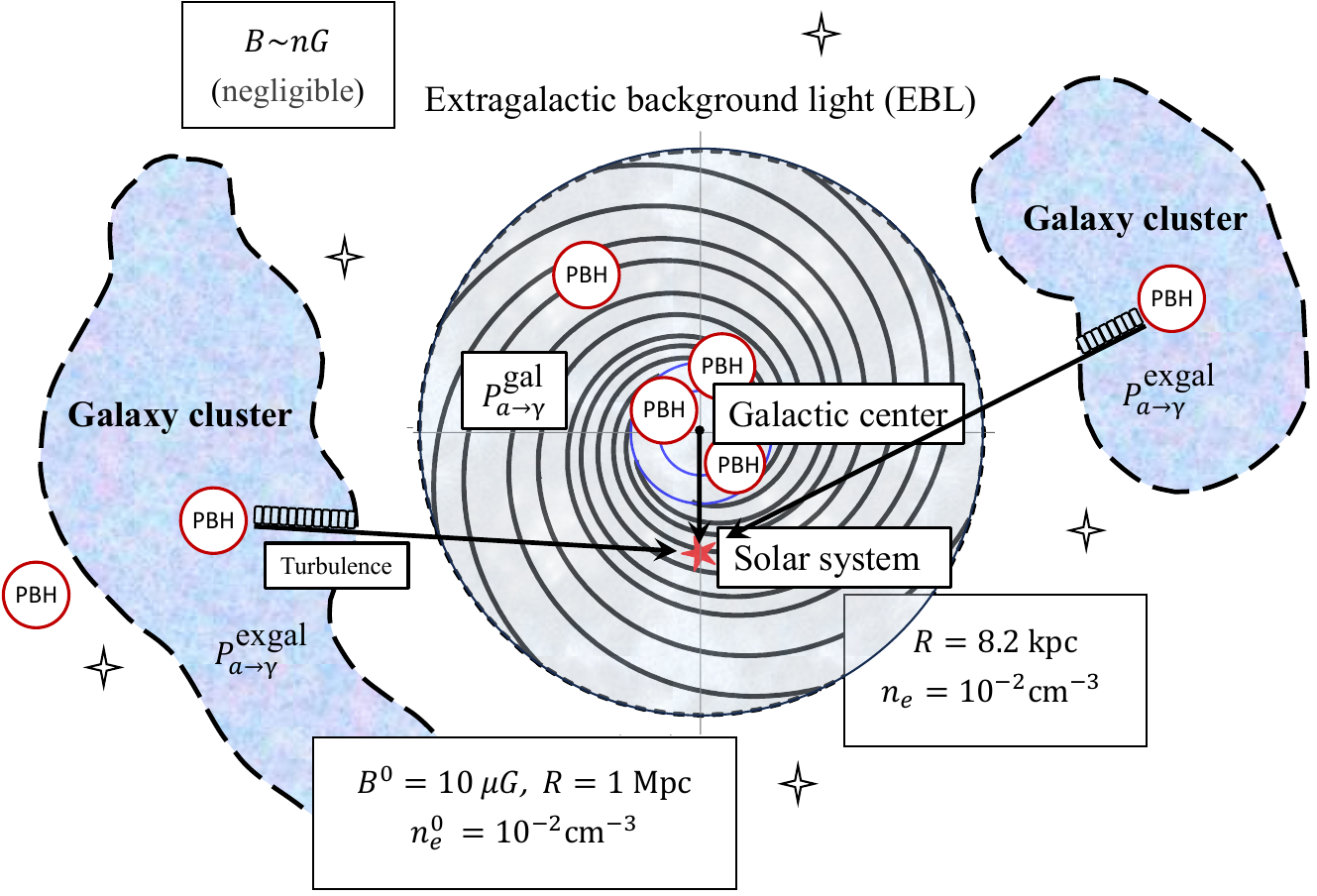}
    \caption{A schematic illustration of conversion from PBH ALPs in cosmic background magnetic fields. }
    \label{fig:diagram}
\end{figure}

The gamma-rays from ALP conversion then yields the following energy spectrum
\begin{equation}
\frac{d^2\Phi_{\rm sig}}{dEd\Omega}=\frac{d^2\Phi_{\rm gal}}{dEd\Omega}\times P_{a\to \gamma}^{\rm gal}(m_a,g_{a\gamma\gamma})+\frac{d^2\Phi_{\rm exgal}}{dEd\Omega}\times P_{a\to \gamma}^{\rm exgal}(m_a,g_{a\gamma\gamma})\;,
\end{equation}
where the superscripts ``gal'' and ``exgal'' of $P_{a\to \gamma}$ represent different regions in the Universe.
In Fig.~\ref{fig:diagram}, we show the schematic illustration of conversion from PBH ALPs in cosmic background magnetic fields. We categorize extragalactic regions into the extragalactic background light (EBL) and galaxy clusters. For the Milky Way and these extragalactic regions, the relavant background parameters are summarized in individual boxes. In this case, we consider only two photon production modes: (1) conversion of PBH ALPs to photons within the Milky Way's magnetic field, and (2) conversion of extragalactic PBH ALPs in galaxy cluster magnetic fields. Due to the weak magnetic field strength of the EBL $B\sim{\rm nG}$~\cite{Planck:2015zrl,Pshirkov:2015tua}, we omit the contribution of PBH particles to the photon signal when passing through this region in subsequent calculations.

\begin{figure}[ht]
\begin{center}
\includegraphics[scale=1,width=0.49\linewidth]{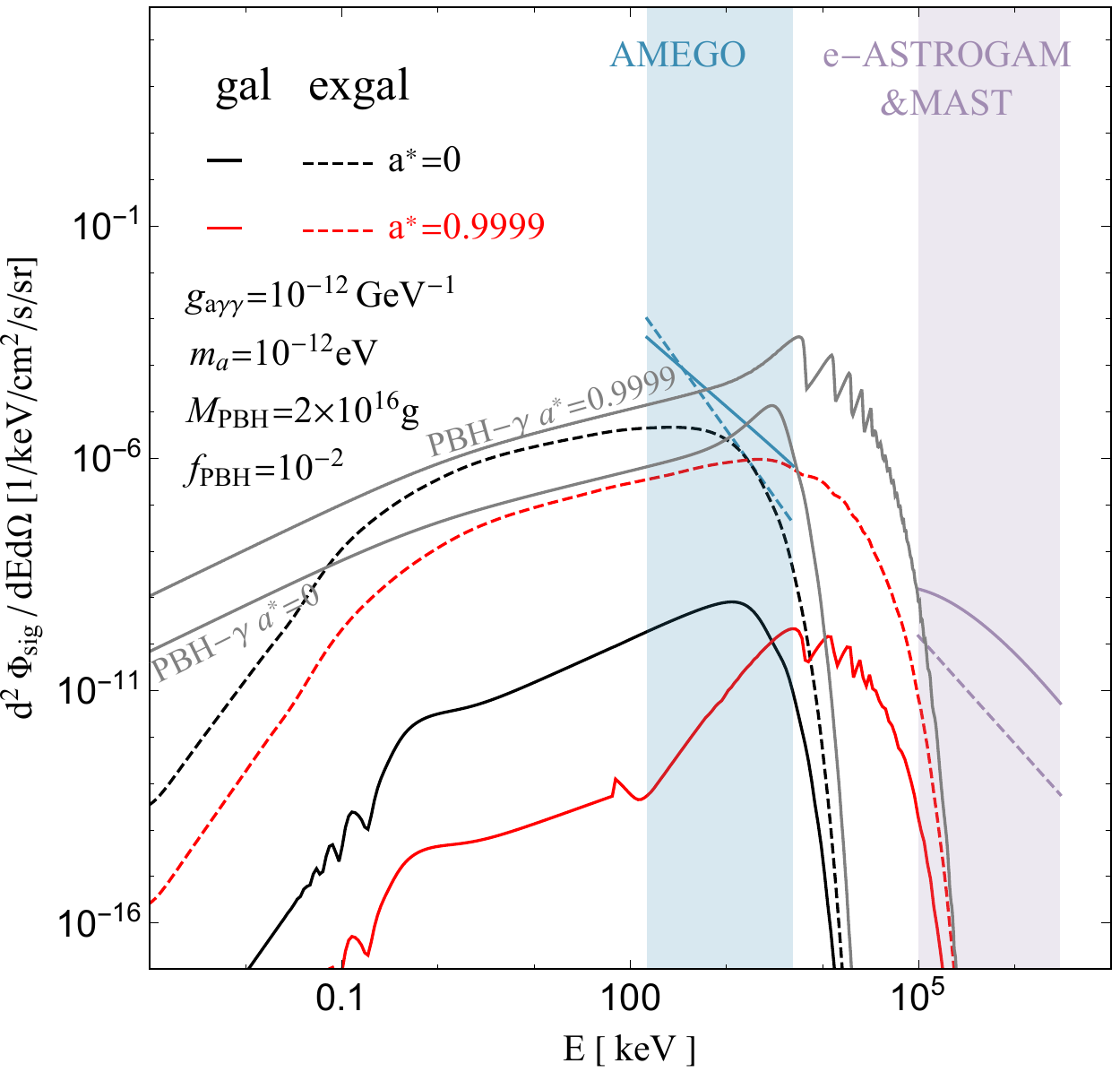}
\includegraphics[scale=1,width=0.49\linewidth]{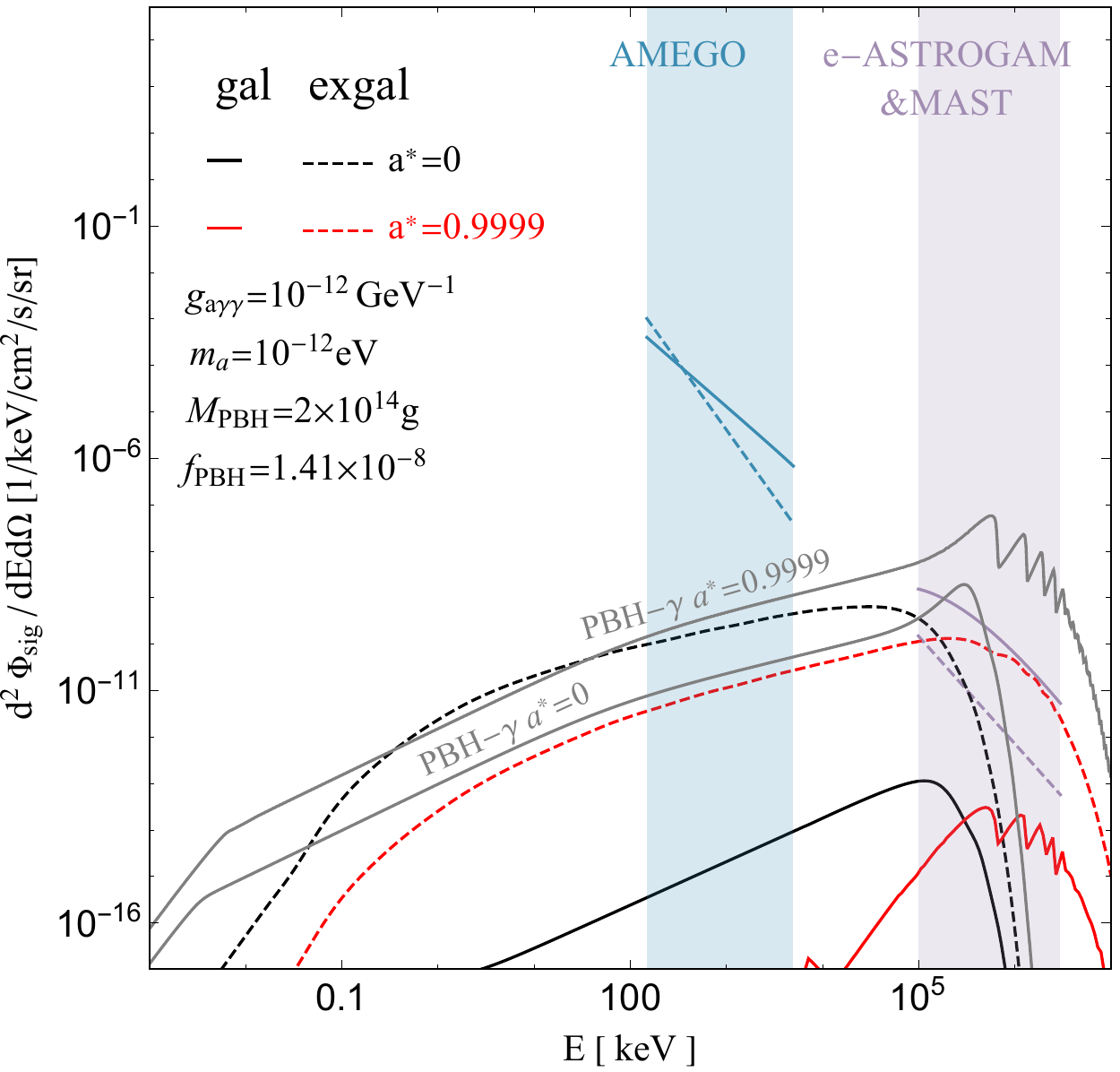}
\end{center}
\caption{
Differential gamma-ray flux from PBH emitted ALPs after propagating through the galactic (solid lines) or extragalactic (dashed lines) background magnetic fields, with $M_{\rm PBH}=2\times10^{16}~\rm g$ and $f_{\rm PBH}=10^{-2}$ (left) or $M_{{\rm PBH}}=2\times10^{14}$ g and $f_{\rm PBH}=1.41\times 10^{-8}$ (right). Two cases of PBH spin with $a^*=0$ (black) and $a^*=0.9999$ (red) are considered and $g_{a\gamma\gamma}$ is fixed to be $10^{-12}~{\rm GeV}^{-1}$. The energy windows are also shown for AMEGO ($150~{\rm keV}-5~{\rm MeV}$, light blue) and e-ASROGAM/MAST ($100~{\rm MeV}-3~{\rm GeV}$, light purple). The total PBH gamma-ray backgrounds are also shown by gray curves.
}
\label{fig:spectrum}
\end{figure}

For this analysis, we focus exclusively on gamma-ray signals within a specific angular region centered on the Milky Way's Galactic Center, defined in Galactic coordinates as $|l|\leq 5^\circ$ and $|b|\leq 5^\circ$.
Fig.~\ref{fig:spectrum} displays the differential gamma-ray flux produced by PBH emitted ALPs after propagating through the galactic (solid lines) or extragalactic (dashed lines) background magnetic fields.
In the left panel of Fig.~\ref{fig:spectrum}, we plot $d^2\Phi_{\rm sig}/dEd\Omega$ for a PBH mass $M_{\rm PBH}=2\times10^{16}~\rm g$ and a fraction $f_{\rm PBH}=10^{-2}$, values permitted by extragalactic gamma-ray constraints~\cite{Carr:2020gox}. Here, we examine two PBH spin scenarios: $a^*=0$ (black) and $a^*=0.9999$ (red), while fixing $g_{a\gamma\gamma}=10^{-12}~{\rm GeV}^{-1}$. The energy windows are also shown for AMEGO ($150~{\rm keV}-5~{\rm MeV}$, light blue) and e-ASROGAM/MAST ($100~{\rm MeV}-3~{\rm GeV}$, light purple).
One can see that the energy spectrum peaks at $E\simeq 1$ MeV, mainly falling within the AMEGO energy window. For the same PBH mass, a larger spin shifts the emission to higher energies, reaching several tens of MeV or even beyond. This is a consequence of angular momentum-enhanced Hawking evaporation. In contrast, spinless PBHs emit predominantly at lower energies, yet their signal flux exceeds that of maximally spinning PBHs ($a^\ast=0.9999$) by an order of magnitude.
In the right panel of Fig.~\ref{fig:spectrum}, we present the signal flux for $M_{{\rm PBH}}=2\times10^{14}$ g and $f_{\rm PBH}=1.41\times 10^{-8}$. Compared with the case in the left panel, the flux peak shifts to higher energies with decreasing PBH mass. Consequently, the e-ASTROGAM and MAST experiments with their higher-energy windows are better suited to detect such signals from lower-mass PBHs.

In addition to the gamma-rays produced through the evaporation-conversion process, there exist other backgrounds in the Universe. For the extragalactic photon background (EGB), current gamma-ray experiments such as Fermi-LAT~\cite{Fermi-LAT:2014ryh}, COMPTEL~\cite{Weidenspointner:2000aq}, and INTEGRAL~\cite{Tsygankov:2007jf} have provided a complete energy spectrum in the range of $\keV-\TeV$. The background flux of extragalactic gamma-ray can be given as
\begin{eqnarray}
\begin{split}
\frac{d^2\Phi_{{\rm exgal}}^{{\rm bkg}}}{dEd\Omega}\Bigg\vert_{150~\keV\leq E\leq 5~\MeV}&=A_{{\rm bkg}}^{{\rm exgal}}\left(\frac{E}{1~\MeV}\right)^{-\alpha^{\rm exgal}}\;,\\
\frac{d^2\Phi_{{\rm exgal}}^{{\rm bkg}}}{dEd\Omega}\Bigg\vert_{100~\MeV\leq E\leq 3~\GeV}&=A_{{\rm bkg}}^{{\rm exgal}}\left(\frac{E}{100~\MeV}\right)^{-\alpha^{\rm exgal}}{\rm exp}\left[-\left(\frac{E}{E_c}\right)\right]\;,
\end{split}
\end{eqnarray}
where the power-law model for the low-energy region has $A_{{\rm bkg}}^{\rm exgal}=0.004135~\MeV^{-1}~{\rm cm}^{-2}~{\rm s}^{-1}~{\rm sr}^{-1}$ and $~\alpha^{{\rm exgal}}=2.8956$~\cite{Ballesteros:2019exr}. The power-law model with exponential cutoff for the high-energy region comes from the Fermi-LAT collaboration~\cite{Fermi-LAT:2014ryh}: $A_{{\rm bkg}}^{\rm exgal}=1.48\times 10^{-7}~\MeV^{-1}~{\rm cm}^{-2}~{\rm s}^{-1}~{\rm sr}^{-1}$,$~\alpha^{{\rm exgal}}=2.31$ and $E_{c}=362~\GeV$. For the galactic gamma-ray background, a fitted model~\cite{Bartels:2017dpb} is adopted in the low-energy region. In the high-energy region, we fit the pion-decay data from Ref.~\cite{Bartels:2017dpb} which dominates in this region. They are given as follows
\begin{eqnarray}
\begin{split}
\frac{d^2\Phi_{{\rm gal}}^{{\rm bkg}}}{dEd\Omega}\Bigg\vert_{150~\keV\leq E\leq 5~\MeV}&=A_{{\rm bkg}}^{{\rm gal}}\left(\frac{E}{1~\MeV}\right)^{-\alpha^{\rm gal}}{\rm exp}\left[-\left(\frac{E}{E_c}\right)^{\gamma^{\rm gal}}\right]\;,\\
\frac{d^2\Phi_{{\rm gal}}^{{\rm bkg}}}{dEd\Omega}\Bigg\vert_{100~\MeV\leq E\leq 3~\GeV}&=A_{{\rm bkg}}^{{\rm gal}}\left(\frac{E}{100~\MeV}\right)^{-\alpha^{\rm gal}}{\rm exp}\left[-\left(\frac{E}{E_c}\right)^{\gamma^{\rm gal}}\right]\;,
\end{split}
\end{eqnarray}
with the best-fit values $A_{{\rm bkg}}^{{\rm gal}}=0.013~\MeV^{-1}{\rm cm}^{-2}~{\rm s}^{-1}~{\rm sr}^{-1}$, $\alpha^{{\rm gal}}=1.8$, $E_c=20~\MeV$ and $\gamma^{{\rm gal}}=2$ for low-energy region and $A_{{\rm bkg}}^{{\rm gal}}=0.00538~\MeV^{-1}{\rm cm}^{-2}~{\rm s}^{-1}~{\rm sr}^{-1}$, $\alpha^{{\rm gal}}=3.32$, $E_c=45708~\MeV$ and $\gamma^{{\rm gal}}=-0.343$ for high-energy region.

In addition, PBHs can emit both ALPs and photons via Hawking evaporation. Ref.~\cite{Ray:2021mxu} explicitly addresses the detectability of gamma-rays from PBH evaporation. For the energy region we consider, the PBH photon flux can be more significant than the observed astrophysical gamma-ray background and thus it plays as the other major background in our case.
We again use \texttt{BlackHawk} to generate the number density of photons from PBH evaporation and show the total fluxes of PBH gamma-ray background using gray curves in Fig.~\ref{fig:spectrum}.

\subsection{Scattering of PBH emitted electron with ALP DM halo}

\begin{figure}[htbp]
\begin{center}
\includegraphics[scale=1,width=0.49\linewidth]{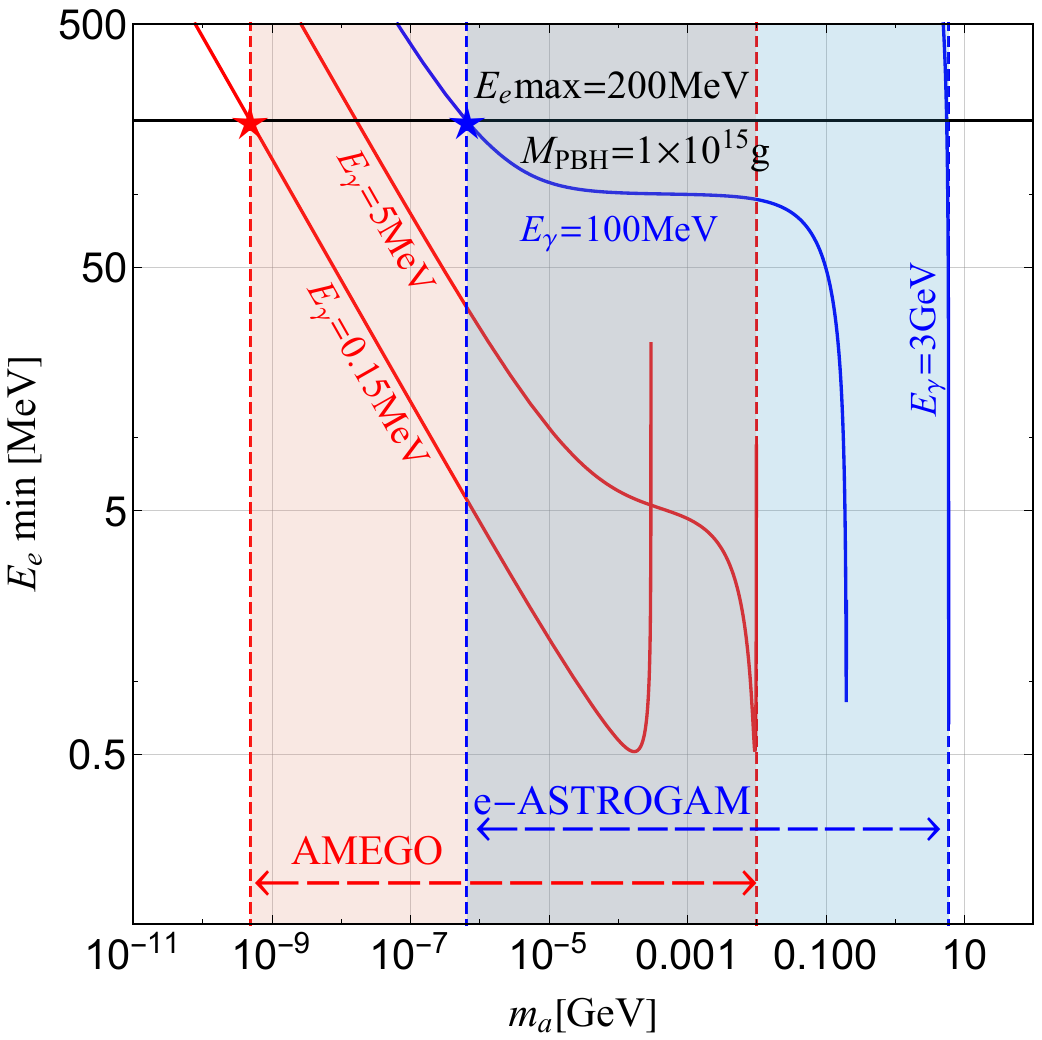}
\includegraphics[scale=1,width=0.49\linewidth]{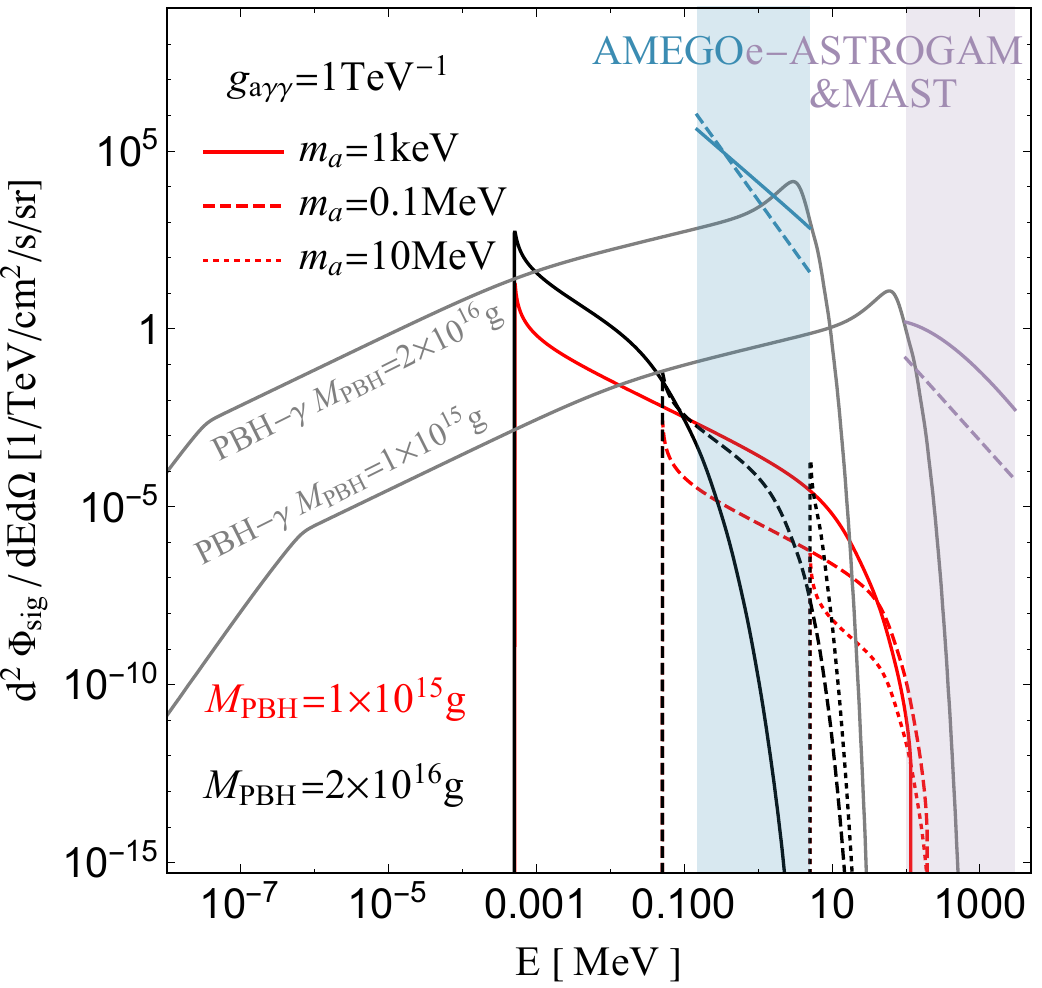}
\end{center}
\caption{Left: The minimum $E_e$ as a function of $m_a$ with $E_\gamma$ being equal to the energy boundaries of AMEGO (red lines) or e-ASTROGAM/MAST (blue lines). We choose $M_{\rm PBH}=1\times 10^{15}$ g for illustration. Right: Gamma-ray spectrum $d^2\Phi_{\rm sig}/dEd\Omega$ from PBH electron scattering off ALP DM with two selected PBH masses $M_{\rm PBH}=1\times10^{15}$ g (red) and $2\times 10^{16}$ g (black), and three ALP masses $m_a=1~\keV$ (solid line), $0.1~\MeV$ (dashed line) and $10~\MeV$ (dotted line). The total PBH gamma-ray backgrounds are also shown by gray curves.
}
\label{fig:spectrum_fermion}
\end{figure}

Next, we consider the scattering between PBH emitted electrons and ALPs in the Galactic DM halo.
In this scenario, DM comprises two components: PBHs and ALPs with their fractional abundances obeying $f_{{\rm PBH}} + f_a = 1$.
Here, the ALP DM acts as a stationary target in the laboratory frame, while PBH electrons with a certain energy scatter off the DM halo. After neglecting the energy losses of PBH electrons during propagation within and beyond the Milky Way, the scattering process produces a real photon in final states through the exchange of a virtual photon in t channel~\footnote{Charged particle such as proton can also generate gamma-ray through the same scattering process~\cite{Goncalves:2025nij}. In this work, we focus exclusively on the scattering of relativistic electron off DM halo.}
\begin{eqnarray}
e^- + a \to e^- + \gamma\;,
\end{eqnarray}
where we assume photophilic ALP with only $g_{a\gamma\gamma}$ coupling.
The gamma-ray energy spectrum is given after integrating over all possible incident PBH electron energies $E_e$~\cite{Dent:2020qev,Goncalves:2025nij}
\begin{eqnarray}
\frac{d^2\Phi_{{\rm sig}}}{dEd\Omega}\Bigg\vert_{E=E_\gamma}=\frac{f_a}{m_a}\int_{\rm RoI}\frac{d\Omega_s}{\Delta\Omega} ds\rho[r(s,l,b)]\int_{E_e^{\rm min}}^{E_e^{\rm max}}dE_e\frac{d\Phi_{e}}{dE_e}\frac{d\sigma_{ea\to e\gamma}}{dE_\gamma}(E_e,E_\gamma)\;.
\label{equ:spectrum}
\end{eqnarray}
Since this evaporation-scattering process requires integrating the PBH emission spectrum over the entire energy range of $E_e$, the energy spectra for $a^*=0$ and $a^*=0.9999$ are nearly identical. Thus, in the following discussion, we assume PBHs to be spinless ($a^*=0$).
For a fixed incoming electron energy, the minimum and maximum gamma-ray energy can be given by kinematic condition~\cite{Dent:2020qev}
\begin{eqnarray}
    E_\gamma^{\rm min/max}=\frac{m_a^2+2m_a E_e}{2E_e+2m_a\pm2\sqrt{E_e^2-m_e^2}}\;,
    \label{equ:kine_condition}
\end{eqnarray}
which yields $E_\gamma^{\rm min}=m_a/2$ and $E_\gamma^{\rm max}\simeq E_e$ at high energies. The minimum $E_e$ as a function of $E_\gamma$ can be obtained by inversely solving the above kinematic condition
\begin{eqnarray}
&E_e^{\rm min}=\Big[-2E_\gamma^2m_a+3E_\gamma m_a^2-m_a^3
+E_\gamma\cos\theta(4E_\gamma^2m_a^2-4E_\gamma m_a^3+m_a^4-2E_\gamma^2m_e^2\\&+8E_\gamma m_a m_e^2-4m_a^2 m_e^2+2E_\gamma^2 m_e^2\cos{2\theta})^{1/2}\Big]\Big/\Big[2(E_\gamma^2-2E_\gamma m_a+m_a^2-E_\gamma^2\cos^2\theta)\Big]\;,
\end{eqnarray}
where the scattering angle $\theta=\pi$ ensure that the minimal energy of incoming electron is sufficient to cause this process.
We set the maximum $E_e$ to achieve 99.99\% integration efficiency for the electron emission flux, e.g., $E_e^{\rm max}=200$ (20) MeV for $M_{\rm PBH}=1\times 10^{15}$ ($2\times 10^{16}$) g. For $M_{\rm PBH}=1\times 10^{15}$ g as illustration, the left panel of Fig.~\ref{fig:spectrum_fermion} shows the minimum $E_e$ as a function of $m_a$ with $E_\gamma$ being equal to the energy boundaries of AMEGO (red lines) or e-ASTROGAM/MAST (blue lines). As a result, we find the certain detectable range of ALP mass for each experiment.

The right panel of Fig.~\ref{fig:spectrum_fermion} illustrates the gamma-ray spectrum $d^2\Phi_{\rm sig}/dEd\Omega$ with two selected PBH masses $M_{\rm PBH}=1\times10^{15}$ g (red) and $2\times 10^{16}$ g (black). The correpsonding values of PBH fraction are allowed by extragalactic gamma-ray constraints: $f_{\rm PBH}=4.22\times 10^{-7}$ and $f_{\rm PBH}=0.01$, respectively. In addition, we also consider three ALP masses $m_a=1~\keV$ (solid line), $0.1~\MeV$ (dashed line) and $10~\MeV$ (dotted line). One can see that the minimum photon energy $E_\gamma^{\rm min}$ shifts to lower values as the ALP mass decreases, as required by the kinematic condition in Eq.~\eqref{equ:kine_condition}. Meanwhile, since the sepctrum Eq.~\eqref{equ:spectrum} is inversely proportional to $m_a$, lighter ALPs are expected to emit more photons under identical conditions.
In addition, lighter PBH generates more enegetic gamma-ray and yields a wider energy range.

\section{Prospective detection of PBH and ALP-photon coupling}
\label{sec:Sensitivity}

Given the energy spectra of the gamma-ray signal and background, we employ the Fisher forecasting method~\cite{Edwards:2017mnf} to numerically evaluate the detection prospects for PBHs and light ALPs. Relative to the Maximum Likelihood Ratio (MLR) method, Fisher forecasting provieds the advantage of requiring fewer parameters and enabling simpler model. The core principle of this approach involves using the Fisher information matrix (computed from the first and second derivatives of the signal spectrum function) to quantify the sensitivity of observed data to selected parameters. This method demonstrates excellent local approximation properties. Mathematically, the Fisher information matrix is an n-dimensional symmetric positive-definite matrix, defined as the second derivative of the log-likelihood function $\ln{\mathcal{L}(\vec\theta\vert\mathcal{D})}$ with respect to n-dimensional parameter vector $\vec{\theta}$ given data set $\mathcal{D}$.
The eigenvalues of the Fisher information matrix quantify the sensitivity of parameter estimation along orthogonal directions in parameter space. 
Ref.~\cite{Bartels:2017dpb} presents a generalized signal-to-noise ratio (SNR) formulation of the Fisher matrix
\begin{eqnarray}
\mathcal{I}_{ij}=\int_{E_{{\rm min}}}^{E_{\rm max}}dE\int_{{\rm ROI}}d\Omega\mathcal{E}(E)\frac{\partial_i\phi(E,\Omega)\partial_j\phi(E,\Omega)}{\phi(E,\Omega\vert\vec\theta=1)}\;,
\end{eqnarray}
where $\vec{\theta}=1$ denotes that all parameters in  $\vec{\theta}$ are set to 1 in the function $\phi(E,\Omega)$, $\phi(E,\Omega)=d^2\Phi/dEd\Omega$ represents the combined contribution of both signal and background fluxes with $\phi=\phi_{{\rm sig}}+\phi_{{\rm bkg}}$, and $\mathcal{E}(E)=T_{{\rm obs}}A_{{\rm eff}}(E)$ denotes the instrumental exposure for gamma-ray detection, combining observation time $T_{{\rm obs}}$ with energy-dependent effective area $A_{{\rm eff}}(E)$. The background flux includes both the observed astrophysical gamma-ray background and the gamma-ray background from PBH evaporation. In this work, we uniformly set the observation time to one year and consider the effective areas of three detectors AMEGO~\cite{Kierans:2020otl}, e-ASTROGAM~\cite{e-ASTROGAM:2016bph} and MAST~\cite{Dzhatdoev:2019kay}.

The Fisher information matrix is used to calculate the reachable limits on a specific set of parameters $\vec\theta$.
In practice, nuisance parameters such as background parameters are typically profiled out to derive the ``profiled Fisher information matrix''. For instance, to project constraints onto the $f_{\rm{PBH}}$ and $M_{\rm PBH}$ parameter space, our model incorporates two types of parameters for a fixed $M_{\rm PBH}$: $f_{\rm PBH}$ as the only parameter of interest (PoI), and the nuisance parameters $A_{\rm{bkg}}^{\rm{exgal}}$, $\alpha^{\rm exgal}$, $A_{\rm{bkg}}^{\rm gal}$, $\alpha^{\rm gal}$, $E_c$ and $\gamma^{\rm gal}$ for AMEGO (or $A_{\rm{bkg}}^{\rm gal}$, $\alpha^{\rm gal}$, $E_c$ and $\gamma^{\rm gal}$ for e-ASTROGAM and MAST). Since the extragalactic background is at least one order of magnitude smaller than the galactic background in the high-energy region, we neglect extragalactic components in the Fisher forecasting for e-ASTROGAM and MAST. The Fisher matrix is partitioned into a block form as follows
\begin{eqnarray}
    \mathcal{I}_{ij}=
    \begin{pmatrix}
        \mathcal{I}_A & \mathcal{I}_C^T \\
        \mathcal{I}_C & \mathcal{I}_B
    \end{pmatrix}\;,
\end{eqnarray}
where the Fisher matrix element $\mathcal{I}_A=\mathcal{I}_{00}$ quantifies the sensitivity of the signal to variations in the PBH fraction $f_{{\rm PBH}}$, and the remaining components, $\mathcal{I}_C$, $\mathcal{I}_C^T$, and $\mathcal{I}_B$, are relevant for background parameters. Using the Schur complement lemma, one can define the profiled Fisher information matrix as $\widetilde{\mathcal{I}}_A=\mathcal{I}_A-\mathcal{I}_C^T~\mathcal{I}_B^{-1}\mathcal{I}_C$. Thus, $\widetilde{\mathcal{I}}_A$ preserves the statistical information of the PoI while eliminating covariance interference introduced by nuisance parameters. The inverse of each diagonal element of ${\widetilde{\mathcal{I}}_A}$ gives the upper limit for the corresponding parameter~\cite{Bartels:2017dpb}
\begin{eqnarray}
\theta_1^{\mathrm{UL}} = f_{\mathrm{PBH}}^{\mathrm{UL}} = Z(p) \sqrt{\widetilde{\mathcal{I}}^{-1}_{A,11}}\;,
\end{eqnarray}
where $Z(p)$ is the cumulative distribution function. We consider the upper limit of $f_{\rm PBH}$ at the 95\% confidence level (C.L.) with $Z(0.05)=1.645$.

\subsection{Prospects from PBH ALP-photon oscillation}

\begin{figure}[htbp]
\begin{center}
\includegraphics[scale=1,width=0.49\linewidth]{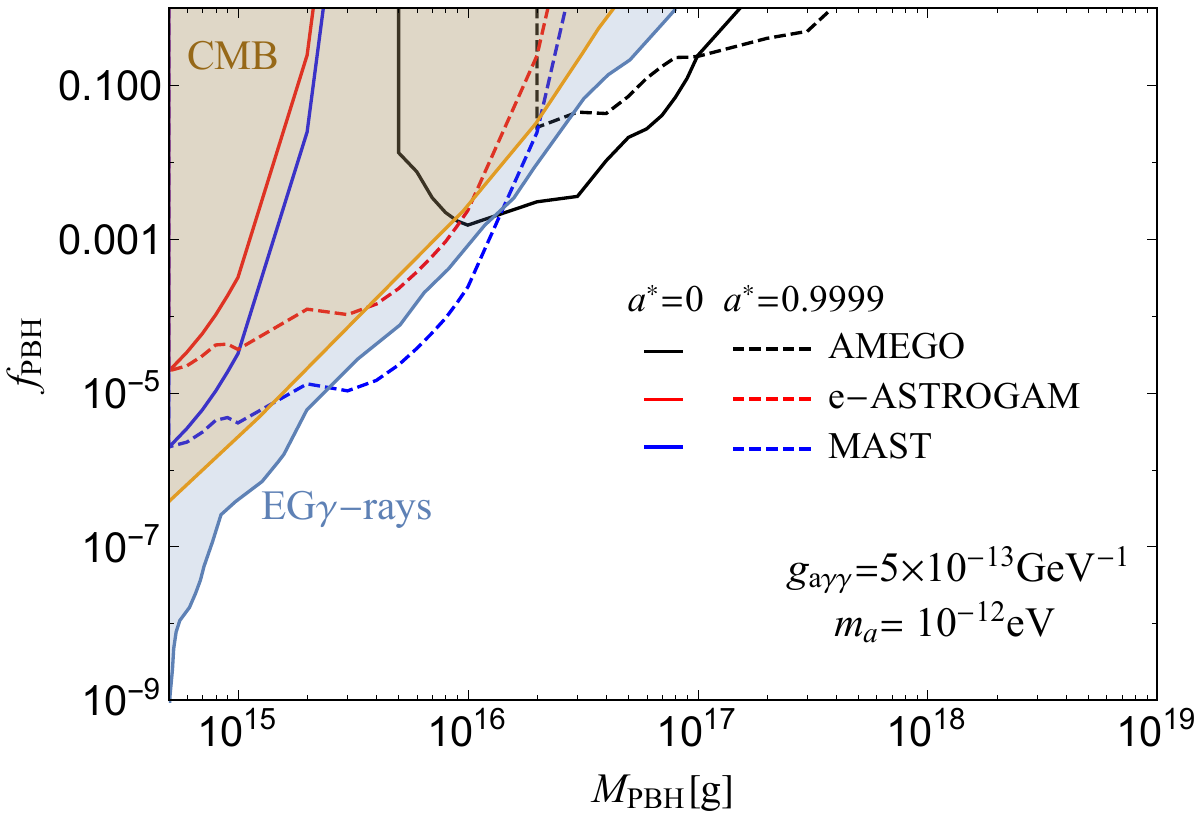}
\includegraphics[scale=1,width=0.49\linewidth]{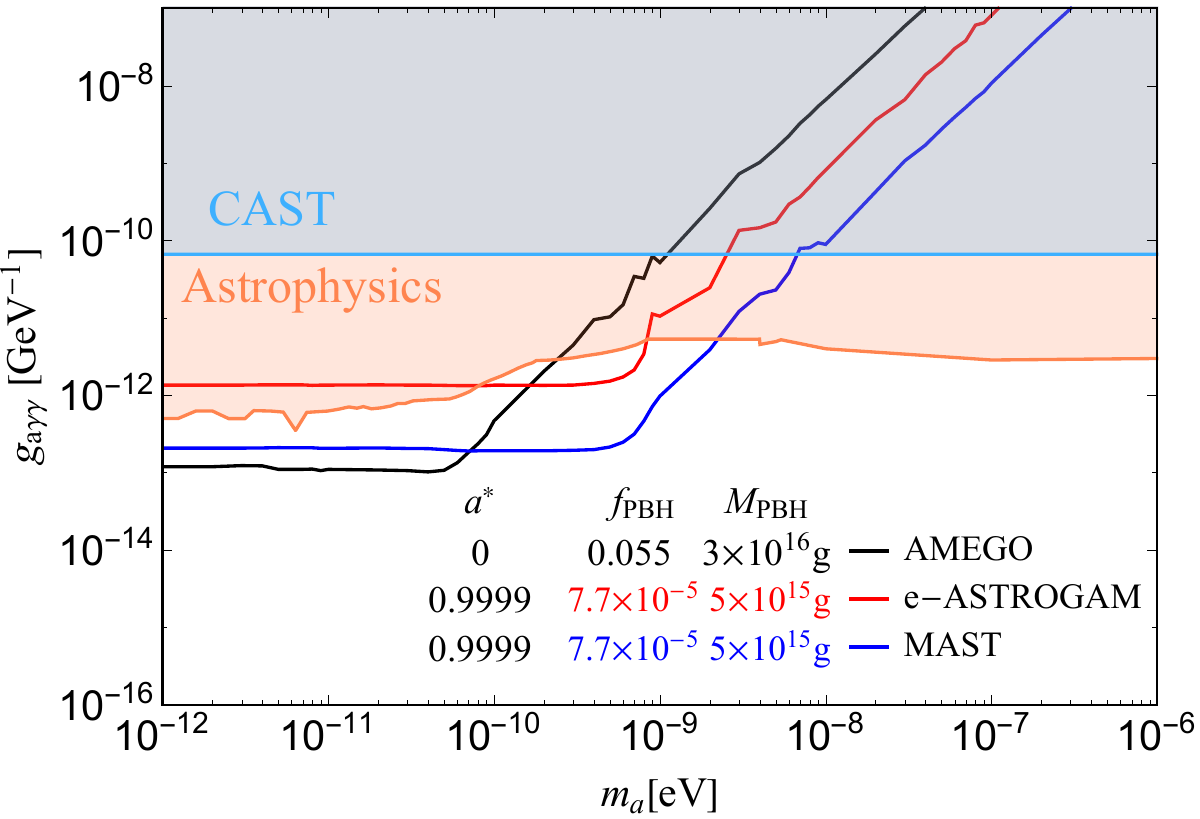}
\end{center}
\caption{Left: Projected 95\% C.L. upper limits on the PBH DM fraction $f_{\rm PBH}$ for the evaporation-conversion scenario from three experiments, with fixed ALP-photon coupling at $g_{a\gamma\gamma} = 5\times10^{-13}~\GeV^{-1}$, ALP mass at $m_a = 10^{-12}~\eV$ and $a^\ast=0$ (solid lines) or $a^\ast=0.9999$ (dashed lines).
The shaded regions are excluded by astrophysical observations~\cite{Carr:2020gox}.
Right: Projected 95\% C.L. upper limits on ALP-photon coupling $g_{a\gamma\gamma}$ from three experiments, with $a^\ast=0$ or $0.9999$ and illustrative $M_{\rm PBH}$ and $f_{\rm PBH}$. The shaded regions show parameter space excluded by CAST and astrophysical observations~\cite{AxionLimits}.
}
\label{fig:upper_limit}
\end{figure}

Based on current null results from gamma-ray observations, the left panel of Fig.~\ref{fig:upper_limit} displays the projected 95\% C.L. upper limits on the PBH DM fraction $f_{\rm PBH}$ for the evaporation-conversion scenario.
We fix the ALP-photon coupling at $g_{a\gamma\gamma} = 5\times10^{-13}~\GeV^{-1}$ and the ALP mass at $m_a = 10^{-12}~\eV$ allowed by ALP constraints.
Solid lines represent limits for spinless PBHs ($a^*=0$), while dashed lines correspond to maximally rotating PBHs ($a^*=0.9999$). For comparison, we also include other $f_{\rm PBH}$ constraints from extragalactic gamma-rays (EG$\gamma$-rays) and the evaporated PBH background in the early Universe (CMB)~\cite{Carr:2020gox}~\footnote{There exist other constraints from Hawking radiation effect on the intergalactic medium temperature evolution~\cite{Mittal:2021egv,Saha:2024ies,Khan:2025kag}.}.
Since heavier PBHs emit predominantly in low-energy regions, AMEGO with lower detectable energy window can be sensitive to a larger PBH mass range $M_{\rm PBH}\gtrsim 5\times 10^{15} $ g ($M_{\rm PBH}\gtrsim 2\times 10^{16} $ g) with viable parameter space for $a^*=0$ ($a^*=0.9999$).
The e-ASTROGAM and MAST experiments with their extended high-energy coverage can probe much ligher PBHs whose corresponding parameter space has significantly been ruled out by existing constraints.

Next, utilizing the known EG$\gamma$-rays constraint on $f_{\rm PBH}$, we reverse the Fisher forecasting procedure for $f_{\rm PBH}$-$M_{\rm PBH}$ to obtain the reachable limits on ALP-photon coupling $g_{a\gamma\gamma}$ for specific PBH masses and allowed $f_{\rm PBH}$ values.
The right panel of Fig.~\ref{fig:upper_limit} shows the 95\% C.L. upper limits on $g_{a\gamma\gamma}$ versus $m_a$ from three experiments for fixed PBH masses and $a^\ast=0$ or $0.9999$.
We choose two illustrative mass benchmarks: $M_{\rm PBH} = 3\times10^{16}$ g for AMEGO, and $M_{\rm PBH} = 5\times10^{15}$ g for e-ASTROGAM and MAST. The maximal values of allowed $f_{\rm PBH}$ are also taken. Our analysis demonstrates that AMEGO can probe ALP-photon coupling down to $g_{a\gamma\gamma}\sim 1\times 10^{-13}~\GeV^{-1}$ for $M_{\rm PBH}=3\times 10^{16}$ g and $m_a<10^{-10}~\eV$, improving upon current astrophysical constraints by one order of magnitude. The limit becomes weak significanlty for $m_a>10^{-10}~\eV$ due to rapid decline of conversion probability within the integral region.
MAST with larger effecitve area achieves more stringent limits than e-ASTROGAM, providing complementary coverage to AMEGO's results in the ALP mass range $m_a>10^{-10}~\eV$.

\subsection{Prospects from scattering of PBH electron off ALP DM}

Due to the extremely small scattering cross-section in the evaporation-scattering scenario, this channel produces significantly weaker constraints on $f_{\rm PBH}$ compared to the evaporation-conversion process. We therefore focus on the projected sensitivity for $g_{a\gamma\gamma}$ and comparing our results with constraints derived from cosmic-ray electrons scattering off ALP DM in Refs.~\cite{Dent:2020qev,Goncalves:2025nij}.

\begin{figure}[htbp]
\begin{center}
\includegraphics[scale=1,width=0.65\linewidth]{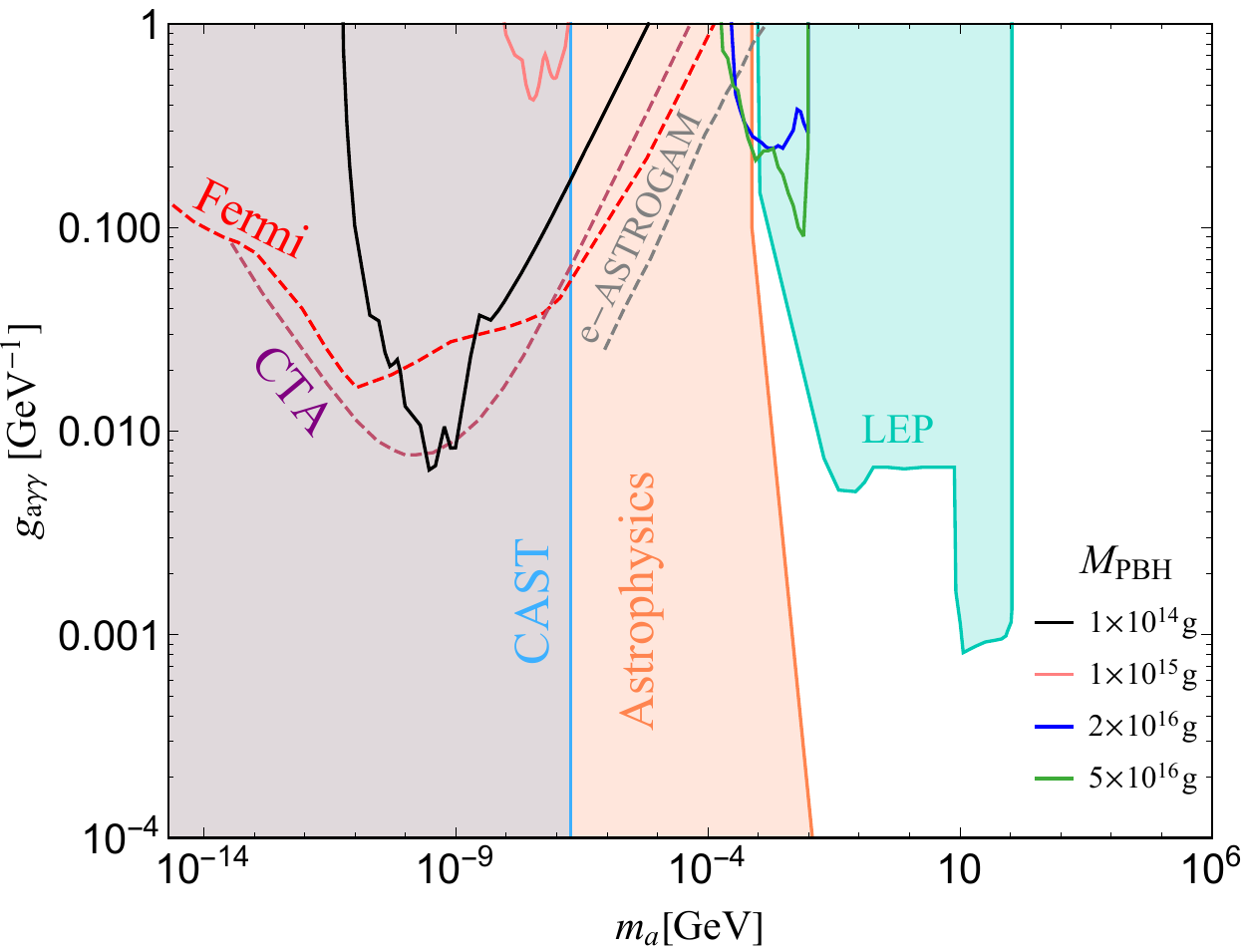}
\end{center}
\caption{Projected limits on $g_{a\gamma\gamma}$ from AMEGO for the evaporation-scattering scenario. We selected four PBHs with different masses and $a^\ast=0$ as electron sources, with their corresponding parameter constraints represented by solid lines: $1\times 10^{14}~{\rm g}$ (black), $1\times 10^{15}~{\rm g}$ (pink), $2\times 10^{16}~{\rm g}$ (blue), and $5\times 10^{16}~{\rm g}$ (green). The constraints on $g_{a\gamma\gamma}$ obtained by cosmic-ray scattering in Fermi, CTA and e-ASTROGAM~\cite{Goncalves:2025nij,Dent:2020qev} are also presented (dashed lines). The shaded regions show parameter space excluded by astrophysical observations and collider experiment~\cite{AxionLimits}.
}
\label{fig:gma_fermion}
\end{figure}

In Fig.~\ref{fig:gma_fermion}, given four PBH mass benchmarks with $a^\ast=0$ as electron source, we present the projected limits on $g_{a\gamma\gamma}$ for AMEGO as illustration.
The results demonstrate that if PBHs as light as $1\times 10^{14}~{\rm g}$ can survive to the present day, the constraint they impose around $m_a\approx 10^{-9}~\GeV^{-1}$ is essentially comparable in magnitude to the strongest limits obtained from cosmic-ray scattering at CTA~\cite{Goncalves:2025nij}.
Furthermore, it turns out that heavier PBHs with $M_{\rm PBH}\gtrsim 10^{16}$ g provide complementary sensitivity reach of $g_{a\gamma\gamma}$ for $m_a\gtrsim 10^{-4}$ GeV. However, due to the constraint on $f_{\rm PBH}$ value, the projected coverage of parameter space has already been excluded by other astrophysical constraint and collider bound.

\section{Conclusion}
\label{sec:Con}

The ALP and PBH are two representative DM candidates as light bosonic DM and macroscopic objects, respectively. In this work, we systematically study the gamma-ray production mechanisms through PBH evaporation and ALP-photon coupling $g_{a\gamma\gamma}$. Furthermore, we evaluate the detection prospects for these signals in next-generation satellite telescopes, including AMEGO, e-ASTROGAM and MAST.

We first propose the evaporation-conversion scenario in which light ALPs are emitted by PBHs and are converted into photons in the presence of magnetic field in the Universe. The second scenario assumes ALPs as dominant DM component in the Milky Way and considers the electron production from PBH evaporation. The emitted electrons scatter off non-relativistic ALP in DM halo and produce gamma-rays through the ALP-photon coupling.
Applying the Fisher forecasting method, we calculate the gamma-ray energy spectra for these two scenarios and derive projected sensitivity for PBH fraction $f_{\rm PBH}$ and ALP-photon coupling $g_{a\gamma\gamma}$. Our main results are summarized as follows.
\begin{itemize}
\item In the evaporation-conversion scenario, AMEGO's enhanced low-energy sensitivity enables probing of an extended PBH mass range ($M_{\rm PBH}\gtrsim 5\times 10^{15}$ g for $a^\ast=0$ and $M_{\rm PBH}\gtrsim 2\times 10^{16}$ g for $a^\ast=0.9999$), covering parameter space still viable under current extragalactic gamma-ray constraints. With their extended
high-energy sensitivity, the e-ASTROGAM and MAST experiments can explore much ligher PBHs whose corresponding parameter space has
significantly been ruled out by existing constraints.
\item AMEGO can also probe ALP-photon coupling down to $g_{a\gamma\gamma}\sim 1\times 10^{-13}~{\rm GeV}^{-1}$ for $m_a<10^{-10}~{\rm eV}$ and heavier PBHs. MAST provides omplementary coverage to
AMEGO’s results in the ALP mass range $m_a>10^{-10}~{\rm eV}$.
\item In the evaporation-scattering scenario, the small scattering cross-section yields significantly weak constraints on $f_{\rm PBH}$. Compared with cosmic-ray electron scattering, heavier PBHs with $M_{\rm PBH}\gtrsim 10^{16}$ g provide complementary sensitivity reach of $g_{a\gamma\gamma}$ for $m_a\gtrsim 10^{-4}$ GeV. However, the projected coverage of parameter space has already been excluded by other astrophysical
constraints or collider bound.
\end{itemize}

\acknowledgments
We would like to thank Wei Chao for useful discussions. T.~L. is supported by the National Natural Science Foundation of China (Grant No. 12375096, 12035008, 11975129).

\bibliography{refs}

\end{document}